\documentclass[pra,twocolumn,superscriptaddress,showpacs,floatfix]{revtex4}
\usepackage{amsmath}
\usepackage{amsthm}
\usepackage{amssymb}
\usepackage{latexsym}
\usepackage{bbm}
\usepackage{graphics}
\theoremstyle{plain}
\newtheorem*{lztheorem}{The Laskowski-$\dot{\text{Z}}$ukowski theorem}
\newtheorem{theorem}{Theorem}
\newtheorem*{lzcorollary}{The Laskowski-$\dot{\text{Z}}$ukowski corollary}

\newtheorem{proposition}{Proposition}
\newtheorem{lemma}{Lemma}
\newtheorem*{observation}{Observation}
\theoremstyle{remark}
\newtheorem*{remark}{Remark}
\newtheorem*{note}{Note}
\theoremstyle{definition}
\newtheorem*{definition}{Definition}
\newcommand{\PP}{\mathrm{P}}
\newcommand{\tr}{\mathrm{Tr}}
\begin{document}
\author{Akira SaiToh}\email{saitoh@qc.ee.es.osaka-u.ac.jp}
\affiliation{Department of Systems Innovation, Graduate School of Engineering
Science, Osaka University, Toyonaka, Osaka 560-8531, Japan}
\author{Robabeh Rahimi}\email{rahimi@math.kindai.ac.jp}
\affiliation{Department of Physics,
Kinki University, Higashiosaka, Osaka 577-8502, Japan}
\title{Population-only decay map for $n$-qubit $n$-partite inseparability
detection}
\date{\today}
%%%Quantum ensemble theory, 05.30.Ch
%%%Quantum entanglement, 03.65.Ud
%%%Quantum information, 03.67.-a
\pacs{05.30.Ch, 03.65.Ud, 03.67.-a}
\begin{abstract}
We introduce a new positive linear map for a single qubit. This map is a
decay only in populations of a single-qubit density operator. It is
shown that an $n$-fold product of this map may be used for a detection of
$n$-partite inseparability of an $n$-qubit density operator ({\em
i.e.}, detection of impossibility of representing a density operator in the
form of a convex combination of products of density operators of
individual qubits). This product map is also investigated in relation to
a variant of the entanglement detection method mentioned by Laskowski and
$\mathrm{\dot{Z}ukowski}$.
\end{abstract}

\maketitle
Detection of entanglement in a mixed state is of public interest.
There have been many methods proposed, such as the partial transpose
criterion \cite{P96}, the range criterion \cite{H97}, the reduction
criterion \cite{H99,C99}, the majorization criterion \cite{NK01}, the
cross norm criterion \cite{R02}, the matrix realignment criterion
\cite{CW03}, the entanglement witness criterion \cite{H96,T01}, etc.
(There are many review articles on these subjects, {\em e.g.},
Refs.\ \cite{HHH01,T02,PV05}.)

Entanglement of a density operator $\rho$ is defined
as inseparability of $\rho$. For a bipartite system consisting of
subsystems $A$ and $B$, $\rho$ is called separable 
iff there is a decomposition \cite{W89}
\begin{equation}\label{eqMS2}
\rho=\sum_i p_i\rho_i^{[A]}\otimes\rho_i^{[B]}
\end{equation}
using non-negative probabilities $p_i$ ($\sum_ip_i=1$). Here, variable $i$
is a label. When this decomposition is impossible, $\rho$ is called
inseparable.

A \textit{multiseparable} \cite{T99} state (or \textit{fully-separable}
\cite{DC00} state) of $n$ qubits is represented by a density operator that
can be decomposed in the form of
\begin{equation}\label{eqMS}
\rho=\sum_ip_i\rho_i^{[1]}\otimes\cdots\otimes\rho_i^{[n]}
\end{equation}
using non-negative probabilities $p_i$ ($\sum_i p_i=1$). Any density
operator that is not multiseparable is called an \textit{$n$-partite
inseparable} density operator of $n$ qubits or an
\textit{$n$-inseparable} density operator of $n$ qubits. We use the term
\textit{$n$-qubit $n$-partite inseparability} in this meaning.

To investigate the inseparability, positive but not completely positive
(PnCP) linear maps are useful \cite{H96,H99,BZ06}. A positive linear map
$\Lambda$ is a map such that
$\sigma\ge0\Longrightarrow\Lambda(\sigma)\ge0$ for a density operator
$\sigma$ ($\sigma\ge0$ means $\langle v|\sigma|v\rangle\ge0$ for $\forall
|v\rangle$ in the Hilbert space $\mathcal{H}$.)
For a bipartite system whose Hilbert space is 
$\mathcal{H}_A\otimes\mathcal{H}_B$, if we have a density operator
$\rho$ in the form of Eq.\ (\ref{eqMS2}), 
$(I^{[A]}\otimes\Lambda^{[B]})(\rho)=\sum_ip_i\rho_i^{[A]}\otimes
\Lambda(\rho_i^{[B]})$ is kept positive. By the contraposition of this
relation, if $(I^{[A]}\otimes\Lambda^{[B]})(\rho)$ is non-positive,
then $\rho$ is inseparable \cite{H96}. Of course, this is true also
when the map $(\Lambda^{[A]}\otimes I^{[B]})$ is used instead of
$(I^{[A]}\otimes\Lambda^{[B]})$.

A similar argument is possible for an $n$-qubit $n$-partite density
operator. When a density operator $\rho$ is in the form of Eq.\ (\ref{eqMS}),
any positive linear map $\Lambda$ acting on a single qubit preserves the
positivity of $(I\otimes \Lambda)(\rho)$. Here, $I$ is the identity map acting
on the rest of the qubits. When $(I\otimes \Lambda)(\rho)$ is a non-positive
operator, then $\rho$ is not multiseparable.

Among such PnCP maps, there are two well-known maps: one is a transpose
$\Lambda_\text{T}$ of a matrix,
$[\Lambda_\text{T}(\sigma)]_{ij}=\sigma_{ji}$ \cite{P96%,W76};
};
the other one is the map \cite{H99} $\Lambda_\text{H}(\sigma)=
\mathbbm{1}\tr\sigma-\sigma$.
These maps have been used for the study of entanglement in bipartite
systems intensively. PnCP maps have been investigated also
for multipartite systems, {\em e.g.}, the depolarizing map \cite{MM04}.

Related to the general multiseparability criterion, recently, one important
theorem was mentioned by Laskowski and $\mathrm{\dot{Z}ukowski}$
\cite{LZ05}:
\begin{lztheorem}%%%[The Laskowski-$\mathrm{\dot{Z}ukowski}$ theorem]
\label{theorem1}
Given a $k$-partite $k$-separable density operator,
the absolute value of an antidiagonal element is less than or equal to
$(1/2)^k$.
\end{lztheorem}
As they claimed that this is obvious, the proof is very simple:
\begin{proof}
Let us write the Hilbert space of a $k$-partite system as
$\mathcal{H}^{[\PP_1]}\otimes\cdots\otimes\mathcal{H}^{[\PP_k]}$ by using
those for subsystems, $\mathcal{H}^{[\PP_j]}$ ($j=1,\ldots,k$). 
Consider a $k$-separable density operator
$\rho^{[\PP_1,\ldots,\PP_k]}_\text{sep}=
\sum_ip_i\rho_i^{[\PP_1]}\otimes\cdots\otimes\rho_i^{[\PP_k]}$ with
$p_i\ge 0$ such that $\sum_ip_i = 1$. 
Any density operator has antidiagonal elements whose absolute values are
less than or equal to $1/2$. Hence, for $\forall i$,
$\rho_i^{[\PP_1]}\otimes\cdots\otimes\rho_i^{[\PP_k]}$ has antidiagonal
elements less than or equal to $(1/2)^k$. Thus
$\rho^{[\PP_1,\ldots,\PP_k]}_\text{sep}$ has antidiagonal elements
less than or equal to $(1/2)^k$.
\end{proof}
By the contraposition of the relation in this theorem, we have
the corollary:
\begin{lzcorollary}%%%[the Laskowski-$\mathrm{\dot{Z}ukowski}$ corollary]
Given a $k$-partite density operator, if it has an antidiagonal
element whose absolute value is larger than $(1/2)^k$, the density
operator is not $k$-separable.
\end{lzcorollary}
This corollary is useful especially in detecting entanglement of
$n$ qubits under the condition that all possible splittings are 
considered. When we find an antidiagonal element whose absolute value
is larger than $(1/2)^n$, we immediately finish this task.

In this report, detection of $n$-qubit $n$-partite inseparability is
concerned. We first introduce a theorem (a variant of the 
Laskowski-$\mathrm{\dot{Z}ukowski}$ corollary) to detect $n$-qubit
$n$-partite inseparability from general off-diagonal elements,
as a subtopic. Second, we will see the main topic where a new positive
linear map $\Lambda_\mathrm{P}$ is defined as a single qubit operation.
We find that the $n$-fold product $\Lambda_\mathrm{P}^{\otimes n}$ may
be used to detect $n$-qubit $n$-partite inseparability. This product map
works, under some condition, for the class of entanglement that can be
detected by the new theorem. In addition to these topics, we compare the
partial-$\Lambda_\mathrm{P}$ map to the partial transpose and show a few
examples of detecting $n$-qubit $n$-partite inseparability.

\begin{theorem}\label{theorem2}
Suppose that we have an $n$-qubit
$n$-partite density operator $\rho=\rho^{[1,\ldots,n]}$.
Then, $\rho$ is not multiseparable if it has an ($a,b$) off-diagonal element
$c_{ab}$ such that $|c_{ab}|>1/2^{h(a,b)}$, where $h(a,b)$ is the Hamming
distance between $a$ and $b$ when expressed in binary notation.
\end{theorem}
\begin{proof}
Let us write the Hilbert space of an $n$-qubit system as
$\mathcal{H}^{[1]}\otimes\cdots\otimes\mathcal{H}^{[n]}$ by using
those for individual qubits, $\mathcal{H}^{[j]}$ ($j=1,\ldots,n$).
Consider an $n$-qubit $n$-separable density operator
$\rho^{[1,\ldots,n]}_\text{sep}
=\sum_ip_i\rho_i^{[1]}\otimes\cdots\otimes\rho_i^{[n]}$ where $p_i\ge 0$
and $\sum_ip_i=1$.
Any density operator of a single qubit has diagonal elements [the
($0,0$) element and the ($1,1$) element] less than or equal to $1$
and the antidiagonal elements [the ($0,1$) element and the ($1,0$)
element] whose absolute values are less than or equal to $1/2$.
Hence for $\forall i$, $\rho_i^{[1]}\otimes\cdots\otimes\rho_i^{[n]}$
has ($a$,$b$) off-diagonal elements $\tilde{c}_{ab}$ for which
$|\tilde{c}_{ab}|\le (1/2)^{h(a,b)}$ is satisfied. This is easy to notice
when we recall that $h(a,b)$ is the number of different bits when $a$
and $b$ are compared as binary strings. Thus,
$\rho^{[1,\ldots,n]}_\text{sep}$ has ($a,b$) off-diagonal elements
$c_{ab}$ such that $|c_{ab}|\le1/2^{h(a,b)}$.
By the contraposition of this relation, the proof is completed.
\end{proof}

Let us look at the illustrating application of the theorem.
The theorem is useful especially when we know the elements of a
density operator. If we find an $(a,b)$ element $c_{a,b}$ whose absolute
value $>1/2^{h(a,b)}$, we immediately find that entanglement is detected
according to Theorem\ \ref{theorem2}. Indeed, in order to detect that a
density operator is not multiseparable, using the theorems is often more
succinct than using
$I^{[1,\ldots,k-1,k+1,\ldots,n]}\otimes\Lambda^{[k]}$ for
several different values of $k$. Consider the three-qubit density operator 
\begin{equation}\label{eqbstate}
 \rho_b^{[1,2,3]}=\frac{1}{7b+1}\begin{pmatrix}
b&0&0&0&0&b&0&0\\
0&b&0&0&0&0&b&0\\
0&0&b&0&0&0&0&b\\
0&0&0&b&0&0&0&0\\
0&0&0&0&\frac{1+b}{2}&0&0&\frac{\sqrt{1-b^2}}{2}\\
b&0&0&0&0&b&0&0\\
0&b&0&0&0&0&b&0\\
0&0&b&0&\frac{\sqrt{1-b^2}}{2}&0&0&\frac{1+b}{2}
\end{pmatrix}
\end{equation}
($0<b<1$).
It is known that this is positive under a partial transpose
with respect to the first qubit (namely, the left most qubit when
expressed in the binary expression). This matrix was originally
introduced by Horodecki \cite{H97} on the basis of the range criterion
as an inseparable density operator with positive partial transposition
for a $2\times 4$ bipartite system. In the present context, our interest
is to find it not multiseparable. One of off-diagonal elements of
$\rho_b^{[1,2,3]}$ is $\langle 7|\rho_b^{[1,2,3]}|4\rangle=\sqrt{1-b^2}/(14b+2)$,
which is larger than $1/2^{h(7,4)}=1/4$ when
$b<(\sqrt{57}-7)/4\simeq0.137$. Thus we find that $\rho_b^{[1,2,3]}$
is not multiseparable according to Theorem\ \ref{theorem2} if
$b<0.137$. In this example, the theorem is, of course, weaker than the
range criterion with which inseparability is detected for $0<b<1$.

It might be of interest which PnCP map can detect entanglement that
can be detected by the theorem. In the next step, we will see a
positive linear map acting on a single qubit. According to 
Propositions\ 1\ and\ 2 shown later, the $n$-fold product of this map
works for this purpose in either the case where we focus solely on
antidiagonal elements or the case where all the phase factors of individual
elements are equal in the left lower side (lower than diagonal elements)
[or, equivalently, in the right upper side (upper than diagonal
elements)] of a density operator.

As we have mentioned, we introduce a positive linear map acting on a
single qubit.
\begin{definition} The positive linear map $\Lambda_\mathrm{P}$ for a
single qubit is defined as follows:
\begin{equation}
 \sigma\Longrightarrow
\Lambda_\mathrm{P}(\sigma) =  \begin{pmatrix}
\frac{\langle0|\sigma|0\rangle+\langle1|\sigma|1\rangle}{2}&
\langle0|\sigma|1\rangle\\
\langle1|\sigma|0\rangle&
\frac{\langle0|\sigma|0\rangle+\langle1|\sigma|1\rangle}{2}
\end{pmatrix},
\end{equation}
where $\sigma$ is a single-qubit density operator and we use the basis
$\{|0\rangle,|1\rangle\}$.
\end{definition}
This map may be called a population-only decay map because of its definition.
\begin{note}
It is easy to find that $\Lambda_\mathrm{P}$ is a positive map
for a single qubit.
For a density operator
\[\sigma=\begin{pmatrix}
c_{00}&x\\x^*&c_{11}
\end{pmatrix},\]
the relation $|x|\le\sqrt{c_{00}c_{11}}\le\frac{1}{2}$ holds.
$\Lambda_\mathrm{P}$ acting on $\sigma$ changes diagonal elements into
$1/2$ and preserves off-diagonal elements. Therefore it is a positive map
for a single qubit.
\end{note}
\begin{remark}[i]
When we see the effect of $\Lambda_\mathrm{P}$ in the Bloch ball of a
single qubit, $\Lambda_\mathrm{P}$ is a projection onto the $x$-$y$
plane as illustrated in Fig.\ \ref{fig1}.
\begin{figure}[hpt]
\begin{center}
\resizebox{0.3\textwidth}{!}{\includegraphics{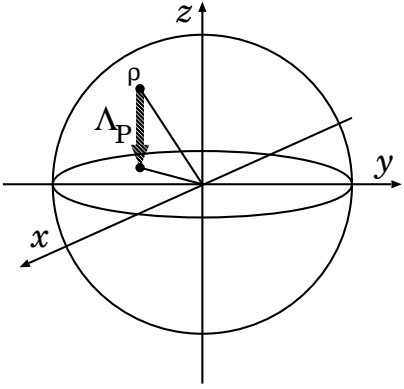}}
\caption{\label{fig1}
The effect (shown as an arrow) of $\Lambda_\mathrm{P}$ acting on a
single-qubit density operator $\rho(x,y,z)=
I/2+x\sigma_X+y\sigma_Y+z\sigma_Z$ illustrated in the Bloch ball
by its coordinate $(x,y,z)$. Here, $\sigma_X,\sigma_Y$, and $\sigma_Z$
are the Pauli matrices. The coordinate is changed to $(x,y,0)$ by
$\Lambda_\mathrm{P}$.}
\end{center}
\end{figure}

\end{remark}
\begin{remark}[ii]
Let us represent the product of the map $\Lambda_\text{P}$ acting on the
$k$th qubit and the identity map acting on the rest of the qubits by
$I^{[1,\ldots,k-1,k+1,\ldots,n]}\otimes\Lambda^{[k]}_\text{P}$. This
product map changes the value of the ($i_1\ldots i_n, j_1\ldots j_n$)
element (here, $i_1,\ldots, i_n, j_1,\ldots, j_n\in \{0,1\}$)
of an Hermitian operator $\xi$ into: \\
(i)
$\frac{1}{2}(\langle i_1\ldots i_{k-1}0i_{k+1}\ldots i_n|\xi|
j_1\ldots j_{k-1}0j_{k+1}\ldots j_n\rangle$\\
$~~~~~+\langle i_1\ldots i_{k-1}1i_{k+1}\ldots i_n|\xi|
j_1\ldots j_{k-1}1j_{k+1}\ldots j_n\rangle)$\\ $~~~~~$when $i_k=j_k$,\\
(ii)
$\langle i_1\ldots i_n|\xi| j_1\ldots j_n\rangle~~~~~~\text{when }i_k\not
=j_k$.\\
Thus, we have
\begin{equation*}\begin{split}
&(I^{[1,\ldots,k-1,k+1,\ldots,n]}\otimes\Lambda_\mathrm{P}^{[k]})(\rho)
=\sum_{\mathbf{a,c,b,d}}\bigl[\\&
\frac{1}{2}\langle\mathbf{ac}|\tr_k\rho|\mathbf{bd}\rangle
(|\mathbf{a}0\mathbf{c}\rangle\langle\mathbf{b}0\mathbf{d}|
+|\mathbf{a}1\mathbf{c}\rangle\langle\mathbf{b}1\mathbf{d}|)\\&
+\langle\mathbf{a}0\mathbf{c}|\rho|\mathbf{b}1\mathbf{d}\rangle
|\mathbf{a}0\mathbf{c}\rangle\langle\mathbf{b}1\mathbf{d}|
+\langle\mathbf{a}1\mathbf{c}|\rho|\mathbf{b}0\mathbf{d}\rangle
|\mathbf{a}1\mathbf{c}\rangle\langle\mathbf{b}0\mathbf{d}|
\bigr],
\end{split}\end{equation*}
where $\tr_k\rho$ is the density operator after tracing out the $k$th
qubit; $\mathbf{a}=i_1\ldots i_{k-1}$, $\mathbf{c}=i_{k+1}\ldots i_n$,
$\mathbf{b}=j_1\ldots j_{k-1}$, and $\mathbf{d}=j_{k+1}\ldots j_n$.
\end{remark}

\begin{observation}
Suppose that we have a multiseparable density operator of $n$ qubits,
\begin{equation}\label{eqMSEP}
\rho=\sum_ip_i\rho_i^{[1]}\otimes\cdots\otimes\rho_i^{[n]}.
\end{equation}
Then, $(I^{[1,\ldots,k-1,k+1,\ldots,n]}\otimes\Lambda_\mathrm{P}^{[k]})(\rho)$
is positive. The contraposition of this relation leads to that, for a
density operator $\rho$, if 
$(I^{[1,\ldots,k-1,k+1,\ldots,n]}\otimes\Lambda_\mathrm{P}^{[k]})(\rho)$
is a non-positive operator, then $\rho$ cannot be represented in the form
of Eq.\ (\ref{eqMSEP}) {\em i.e.} $\rho$ is not multiseparable.
Similarly, for a density operator $\rho$, if 
$(\Lambda_\mathrm{P}^{\otimes n})(\rho)$ is a non-positive operator, 
then $\rho$ is not multiseparable.
\end{observation}
As is well known, this logic was explained intensively by M. Horodecki
and P. Horodecki \cite{H99}. Now we will see two remarks relating to this
observation.
\begin{remark}[i]
In detecting entanglement of two qubits, the partial population-only
decay $I\otimes\Lambda_\text{P}$ is not so strong as the partial transpose
$I\otimes\Lambda_\text{T}$, where $\Lambda_\text{T}$ is the single-qubit
transpose operation.
For a two-qubit density operator $\rho^{[1,2]}$, we can write that
\begin{equation}
 (I\otimes\Lambda_\text{P})(\rho^{[1,2]})=
\frac{1}{2}[\rho^{[1,2]}+
(I\otimes\Lambda_{X})(I\otimes\Lambda_\text{T})(\rho^{[1,2]})],
\end{equation}
where $\Lambda_{X}$ is a single-qubit NOT operation:
$\Lambda_{X}(\sigma)=X\sigma X$ 
(here, $\sigma$ is a $2\times2$ matrix and
$X=|0\rangle\langle1|+|1\rangle\langle0|$).
Thus, $(I\otimes\Lambda_\text{P})(\rho^{[1,2]})$ is sometimes positive for
a density operator having the property of negative partial transposition.
It is clear that $(I\otimes\Lambda_\text{T})(\rho^{[1,2]})$ is
non-positive if $(I\otimes\Lambda_\text{P})(\rho^{[1,2]})$ is non-positive.

For example, consider the two-qubit isotropic states
 \cite{H99,B05}, \[\rho_\text{iso}^{[1,2]}=\frac{
|\phi\rangle\langle\phi|+s \mathbbm{1}/4}{1+s},\]
where $|\phi\rangle\in\{
\frac{1}{\sqrt{2}}(|00\rangle\pm |11\rangle),
\frac{1}{\sqrt{2}}(|01\rangle\pm |10\rangle)\}$;
$s\le -4$ or $0\le s$.
It is known that $\rho_\text{iso}^{[1,2]}$ is entangled for
$0\le s <2$ because $(I\otimes\Lambda_\text{T})(\rho_\text{iso}^{[1,2]})$
has eigenvalues $(s-2)/(4s+4)$ and $(s+2)/(4s+4)$ (with the multiplicity
of 3 for the latter one) irrespectively of $|\phi\rangle$. When we use
the partial population-only decay map, we have
\begin{equation*}\begin{split}
(I\otimes\Lambda_\text{P})(\rho_\text{iso}^{[1,2]})\in
 \{&\frac{\mathbbm{1}}{4}\pm\frac{1}{2(1+s)}
(|00\rangle\langle11|+|11\rangle\langle00|),\\
&\frac{\mathbbm{1}}{4}\pm\frac{1}{2(1+s)}
(|01\rangle\langle10|+|10\rangle\langle01|)\}.
\end{split}\end{equation*}
We find that
$(I\otimes\Lambda_\text{P})(\rho_\text{iso}^{[1,2]})$ has eigenvalues
$(s-1)/(4s+4)$, $(s+3)/(4s+4)$, and $1/4$ with the multiplicity of 2
irrespectively of $|\phi\rangle$. The first eigenvalue is negative for
$0\le s < 1$. This interval in which entanglement is detected is half of
that we obtain when $I\otimes\Lambda_\text{T}$ is used.
\end{remark}

\begin{remark}[ii]
It should be remarked that $\Lambda_\text{P}^{\otimes n}$ can be used
for detecting entanglement. In contrast, if we use the single-qubit transpose
operation $\Lambda_\text{T}$, $\Lambda_\text{T}^{\otimes n}$ is nothing
but a transpose and it cannot be used for the detection of entanglement.

For example, consider an obviously
entangled pure state
$|\varphi\rangle=\sqrt{p}|00\rangle+\sqrt{1-p}|11\rangle$ with $0<p<1$.
Then,
\[
 (I\otimes\Lambda_\text{P})(|\varphi\rangle\langle\varphi|)=
\begin{pmatrix}
\frac{p}{2}&0&0&\sqrt{p(1-p)}\\0&\frac{p}{2}&0&0\\0&0&\frac{1-p}{2}&0\\
\sqrt{p(1-p)}&0&0&\frac{1-p}{2}
\end{pmatrix}.
\]
This has four eigenvalues, $p/2,(1-p)/2,1/4\pm(\sqrt{-12p^2+12p+1})/4$.
It has a negative eigenvalue when $0<p<1$.
Similarly, for the same pure state,
\[
(\Lambda_\text{P}\otimes\Lambda_\text{P})(|\varphi\rangle\langle\varphi|)=
\begin{pmatrix}
\frac{1}{4}&0&0&\sqrt{p(1-p)}\\0&\frac{1}{4}&0&0\\0&0&\frac{1}{4}&0\\
\sqrt{p(1-p)}&0&0&\frac{1}{4}
\end{pmatrix}.
\] 
This has four eigenvalues, $1/4,1/4,1/4\pm\sqrt{p-p^2}$.
It has a negative eigenvalue when
$\frac{1}{2}-\frac{\sqrt{3}}{4}<p<\frac{1}{2}+\frac{\sqrt{3}}{4}$.
Indeed, the entanglement detection using $\Lambda_\text{P}\otimes
\Lambda_\text{P}$ does not
work for the intervals $0<p<\frac{1}{2}-\frac{\sqrt{3}}{4}\simeq0.067$ and
$1>p>\frac{1}{2}+\frac{\sqrt{3}}{4}\simeq0.933$, but still works for the
wide range of $p$.
\end{remark}

Using $\Lambda_\mathrm{P}^{\otimes n}$, we prove the following propositions using
lemmas shown later.
\begin{proposition}\label{proposition1}
For an $n$-qubit density operator
$\rho=\rho^{[1,\ldots,n]}$, $(\Lambda_\mathrm{P}^{\otimes n})(\rho)\not\ge0$ holds
if $\rho$ has an ($a,b$) antidiagonal element $c_{ab}$ [{\em i.e.}, $h(a,b)=n$]
such that $|c_{ab}|>1/2^n$ where $h(a,b)$ is the Hamming distance between $a$
and $b$ when expressed in binary notation and $a,b\in\{0,\ldots,2^n-1\}$.
\end{proposition}
\begin{proof}
Consider the $n$-qubit density operator $\rho$. It is
assumed that it has some off-diagonal element $c_{ab}$ %%%%($a\not = b$)
such that $h(a,b)=n$ and $|c_{ab}|> 1/2^n$.
Then apply the operator $\Lambda_\mathrm{P}^{\otimes n}$ to $\rho$.
This leads to that
\[\begin{split}
\sigma=(\Lambda_\mathrm{P}^{\otimes n})(\rho)&=
\mathrm{diag}(1/2^n,\ldots,1/2^n)\\
&~~+\text{off-diagonal elements}.
\end{split}\]
Note that the ($a,b$) element is unchanged by
 $\Lambda_\mathrm{P}^{\otimes n}$ when
$h(a,b)=n$. Thus $|\langle a|\sigma|b\rangle|=|c_{ab}|>1/2^n$.
Hence $\sigma$ is a non-positive operator according to Lemma\ 2.
\end{proof}
\begin{proposition}\label{proposiyion2}
Suppose that we have an $n$-qubit density operator
$\rho=\rho^{[1,\ldots,n]}$ such that its off-diagonal elements $c_{ij}$
satisfy the relation $\mathrm{Arg}(c_{ij})=\mathrm{Arg}(c_{kl})$ for
$i>j$ and  $k>l$. Then, $(\Lambda_\mathrm{P}^{\otimes n})(\rho)\not\ge0$
holds if $\rho$ has an ($a,b$) off-diagonal element $c_{ab}$ such that
$|c_{ab}|>1/2^{h(a,b)}$.
\end{proposition}
\begin{proof}
Consider an $n$-qubit density operator $\rho$ with off-diagonal elements
$c_{ij}$ such that $\mathrm{Arg}(c_{ij})=\mathrm{Arg}(c_{kl})$ for
$i>j$ and $k>l$. It is assumed that it has some off-diagonal element
$c_{ab}$ %%%($a\not = b$) 
such that $|c_{ab}|> 1/2^{h(a,b)}$.
Then apply the operator $\Lambda_\mathrm{P}^{\otimes n}$ to $\rho$.
This leads to
\[\begin{split}
\sigma=(\Lambda_\mathrm{P}^{\otimes n})(\rho)&=
\mathrm{diag}(1/2^n,\ldots,1/2^n)\\
&~~+\text{off-diagonal elements}.
\end{split}\]
According to Lemma\ 1, $|\langle a|\sigma|b\rangle|
\ge |c_{ab}|/2^{n-h(a,b)} > 1/2^n$. Hence $\sigma$ is a non-positive
operator according to Lemma\ 2.
\end{proof}
These propositions illustrate an ability of $\Lambda_\mathrm{P}^{\otimes
n}$: it fails to map a density matrix to a non-positive Hermitian matrix
for some obviously entangled state, as we have seen, but it succeeds for
some state positive under a partial transpose, such as the state of
Eq.\ (\ref{eqbstate}) for $b<0.137$.

In proving above propositions, we have utilized the following lemmas.
\begin{lemma}
Suppose that an $n$-qubit density operator $\rho$ has ($i,j$) elements
$c_{ij}$ such that $\mathrm{Arg}(c_{ij})=\mathrm{Arg}(c_{kl})$
for $i>j$ and $k>l$. Consider the ($a,b$) element $c_{ab}$ of $\rho$.
Then, %%%for the map $\Lambda_\mathrm{P}$ defined above,
$(\Lambda_\mathrm{P}^{\otimes n})(\rho)$ has the ($a,b$) element 
$\tilde{c}_{ab}$ such that $|\tilde{c}_{ab}|\ge |c_{ab}|/2^{n-{h(a,b)}}$.
\end{lemma}
\begin{proof}
Consider binary bits $x$ and $y$; binary strings $\mathbf{a}$ and
$\mathbf{b}$ with bit length $k-1$; binary strings $\mathbf{c}$ and
$\mathbf{d}$ with bit length $n-k$.
Consider the element of $\rho$,
$\langle \mathbf{a}x\mathbf{c}|\rho|\mathbf{b}y\mathbf{d}\rangle$.
Let $\rho$ evolve under
$I^{[1,\ldots,k-1,k+1,\ldots, n]}\otimes\Lambda_\mathrm{P}^{[k]}$. Then,
$\langle \mathbf{a}x\mathbf{c}|
(I^{[1,\ldots,k-1,k+1,\ldots, n]}\otimes\Lambda_\mathrm{P}^{[k]})(\rho)
|\mathbf{b}y\mathbf{d}\rangle$ is equal to 
$\frac{1}{2}(\langle \mathbf{a}0\mathbf{c}|\rho|\mathbf{b}0\mathbf{d}\rangle
+\langle \mathbf{a}1\mathbf{c}|\rho|\mathbf{b}1\mathbf{d}\rangle)$ if
 $x=y$; otherwise it is equal to
$\langle \mathbf{a}x\mathbf{c}|\rho|\mathbf{b}y\mathbf{d}\rangle$.
Note that if $\mathbf{a}0\mathbf{c}>\mathbf{b}0\mathbf{d}$,
then $\mathbf{a}1\mathbf{c}>\mathbf{b}1\mathbf{d}$; if
$\mathbf{a}0\mathbf{c}<\mathbf{b}0\mathbf{d}$, then
$\mathbf{a}1\mathbf{c}<\mathbf{b}1\mathbf{d}$. Hence,
$I^{[1,\ldots,k-1,k+1,\ldots, n]}\otimes\Lambda_\mathrm{P}^{[k]}$ does
not change the argument but changes the absolute value of the
$(\mathbf{a}x\mathbf{c},\mathbf{b}y\mathbf{d})$ element by the factor
$\ge 1/2$ in the case of $x=y$, because of equal argument values in
the $(\mathbf{a}0\mathbf{c},\mathbf{b}0\mathbf{d})$ and 
$(\mathbf{a}1\mathbf{c},\mathbf{b}1\mathbf{d})$ elements.
Thus, if we continue to apply
$I^{[1,\ldots,k-1,k+1,\ldots, n]}\otimes\Lambda_\mathrm{P}^{[k]}$ from
$k=1$ to $k=n$, then for each $k$, reduction in the absolute value by
the factor $\ge1/2$ occurs if $x = y$.
Therefore $|c_{ab}|/2^{n-h(a,b)}$ is a lower bound of $|\tilde{c}_{ab}|$.
\end{proof}
\begin{lemma} Suppose that an Hermitian operator $\sigma$ has an
($a,b$) element $s_{ab}\in\mathbbm{C}$ ($a\not = b$) such that
$|s_{ab}|> 1/2^n$ and has diagonal elements
$s_{ii}=1/2^n$ for $\forall i$. Then, $\sigma$ is a non-positive operator.
\end{lemma}
\begin{proof}
Let us use the vector $|v\rangle=|a\rangle-
(s_{ab}^*/|s_{ab}|)|b\rangle$. Then we have
$\langle v|\sigma|v\rangle=s_{aa}+s_{bb}-|s_{ab}|-|s_{ab}| 
=2(1/2^n-|s_{ab}|)< 0$.
\end{proof}

To summarize, we have shown a variant of the
Laskowski-$\mathrm{\dot{Z}ukowski}$ corollary that can be used to detect the
$n$-qubit $n$-partite inseparability from any one of off-diagonal
elements of a density operator.  To detect the $n$-qubit $n$-partite
inseparability, we have also introduced and investigated a new positive
map $\Lambda_\mathrm{P}$ acting on a single qubit and its $n$-fold
product, $\Lambda_\mathrm{P}^{\otimes n}$.

The authors are thankful to Mikio Nakahara for helpful discussions and 
reading through the manuscript carefully. A.~S. is supported by the JSPS
Research Fellowship for Young Scientists.


\begin{thebibliography}{99}
\bibitem{P96} A.\ Peres, Phys.\ Rev.\ Lett.\ \textbf{77}, 1413 (1996).
\bibitem{H97} P.\ Horodecki, Phys.\ Lett.\ A \textbf{232}, 333 (1997).
\bibitem{H99} M.\ Horodecki and P.\ Horodecki, Phys.\ Rev.\ A \textbf{59},
	4206 (1999).
\bibitem{C99} N.~J.\ Cerf, C.\ Adami, and R.~M.\ Gingrich, Phys.\ Rev.\ A
 \textbf{60}, 898 (1999).
\bibitem{NK01} M.~A.\ Nielsen and J.\ Kempe, Phys.\ Rev.\
	Lett. \textbf{86}, 5184 (2001).
\bibitem{R02} O.\ Rudolph, e-print arXiv:quant-ph/0202121.
\bibitem{CW03} K.\ Chen and L-A\ Wu, Quant.\ Inf.\ Comput. \textbf{3} (3),
193 (2003).
\bibitem{H96} M.\ Horodecki, P.\ Horodecki, and R.\ Horodecki,
Phys.\ Lett.\ A \textbf{223}, 1 (1996).
\bibitem{T01} B.~M.\ Terhal, Phys.\ Lett.\ A \textbf{271}, 319 (2000).
\bibitem{HHH01} M.\ Horodecki, P.\ Horodecki, and R.\ Horodecki, e-print
	arXiv:quant-ph/0109124.
\bibitem{T02} B.~M.\ Terhal, Theor.\ Comput.\ Sci. \textbf{287}, 313
	(2002).
\bibitem{PV05} M.~B.\ Plenio and S.\ Virmani, e-print arXiv:quant-ph/0504163.
\bibitem{W89} R.~F.\ Werner, Phys.\ Rev.\ A \textbf{40}, 4277 (1989).
\bibitem{T99} A.~V.\ Thapliyal, Phys.\ Rev.\ A \textbf{59}, 003336
	(1999).
\bibitem{DC00} W.~D\"ur and J.~I.\ Cirac, Phys.\ Rev.\ A \textbf{61},
	042314 (2000).
\bibitem{BZ06} I.\ Bengtsson and K.\ $\mathrm{\dot{Z}yczkowski}$,
\textit{Geometry of quantum states: An introduction to quantum entanglement}
	(Cambridge University Press, Cambridge, 2006), Chap.\ 15.4.
%\bibitem{W76} S.~L.\ Woronowicz, Rep.\ Math.\ Phys. \textbf{10}, 165 (1976).
\bibitem{MM04} V.~I.\ Man'ko, G.\ Marmo, E.~C.~G.\ Sudarshan, and F.\ Zaccaria,
Phys.\ Lett.\ A \textbf{327}, 353 (2004).
\bibitem{LZ05} W.\ Laskowski and M.\ $\mathrm{\dot{Z}ukowski}$,
Phys.\ Rev.\ A \textbf{72}, 062112 (2005). 
\bibitem{B05} R.~A.\ Bertlmann, K.\ Durstberger, B.~C.\ Hiesmayr, and
	P.\ Krammer, Phys.\ Rev.\ A \textbf{72}, 052331 (2005).

\end{thebibliography}
\end{document}